\documentstyle[12pt,twoside,fleqn,espcrc1]{article}

\newcommand{\AmS}{{\protect\the\textfont2
  A\kern-.1667em\lower.5ex\hbox{M}\kern-.125emS}}

\title{Projective Statistics and Spinors in Hilbert Space}

\author{Frank Wilczek\address{Institute for Advanced Study, 
        School of Natural Science, \\ 
        Olden Lane, Princeton, NJ 08540}
        \thanks{Currently on leave at Leiden University.
        Instituut-Lorentz, Niels Bohrweg 2, 2333 CA Leiden, The
        Netherlands.  Supported in part by DOE grant
        DE-FG02-90ER40542.  IASSNS-HEP98-61}}

\begin{document}

\maketitle

\begin{abstract}

In quantum mechanics, symmetry groups can be realized by projective,
as well as by ordinary unitary, representations.  For the permutation
symmetry relevant to quantum statistics of $N$ indistinguishable
particles, the simplest properly projective representation is highly
non-trivial, of dimension $2^{[{N-1\over 2}]}$, and is most easily
realized starting with spinor geometry.  Quasiparticles in the
Pfaffian quantum Hall state realize this representation.  Projective
statistics is a consistent theoretical possibility in any dimension.

\end{abstract} 

\bigskip

Symmetry operations in quantum mechanics are implemented by unitary
operators,
$g\rightarrow U(g)$. Since observable amplitudes involve products of
bras and kets, an overall phase factor has no effect on observables.
Thus one must consider the possibility that the group multiplication
is realized only ``up to a phase'', or projectively, with

\begin{equation}
U(gh) ~=~ \eta (g, h) U(g) U(h)~.
\label{projective}
\end{equation}

This possibility is realized, famously, for the spinor representation
of rotations.  A rotation around axis $\hat n$ through angle $\theta$
is implemented by $e^{i{\hat n} {\vec \sigma } \theta /2}$.  This is
ambiguous, as the transformations for $\theta$ and $\theta + 2\pi$
differ by a sign.  We can of course remove the ambiguity by
restricting $-\pi < \theta
\leq \pi$, but then there will be non-trivial signs in the
multiplication law~\ref{projective}.
As Weyl emphasized, the basic commutation rules $[p_k , q_l ] = -i
\delta_{kl}$ state that the abelian group of canonical translations is
realized projectively ~\cite{weyl}.

Another very basic quantum mechanical symmetry concerns the
interchange, or permutation, of indistinguishable particles.  It is
natural to ask whether the permutation symmetry is realized
projectively in Nature.  The mathematical theory of projective
representations of the group $S_N$ of permutations of $N$ elementary
particles was developed in classic papers by I. Schur ~\cite{schur}, prior to
the discovery of either modern quantum mechanics or spinors.  The
simplest (irreducible) non-trivial projective representations of
$S_N$ are already surprisingly intricate and have dimensions which grow
exponentially with $N$.  They are intimately related to spinor
representations of $SO(N)$, as I shall now explain.

{}From $\Gamma$ matrices satisfying the Clifford algebra $\{\Gamma_j,
\Gamma_k\} = 2 \delta_{jk}$ one can form commutators $L_{jk} \equiv -
{i\over 4} [\Gamma_j, \Gamma_k ] $ that satisfy the Lie algebra of
$SO(N)$.  $L_{jk}$ is the infinitesimal generator  of rotations in
the jk plane.  Note that rotation through $\pm \pi$ yields
representatives differing in sign.

Conjugation with  $\Gamma_k$  reverses the sign of all the other
$\Gamma$s, so it
implements reflection in the $k$ axis.   Similarly, $(\Gamma_j +
\Gamma_k)/ \sqrt 2$ implements a reflection that interchanges the j
and k axes.  One might therefore expect to realize the permutation
group inside a spinor representation using these elements.
This is close to the truth, but there is a subtlety.

It is not difficult to convince oneself that the abstract
``interchanges'' $T_{1;2}$,
$T_{2;3}$, ... , $T_{N-1;N}$  subject to the first three of the
relations

\begin{eqnarray}
(T_{j;j+1})^2 &=& 0 \nonumber \\
(T_{j;j+1} T_{j+1; j+2})^3 &=& 1 \nonumber \\
T_{j;j+1} T_{k; k+1} &=& + T_{k;k+1} T_{j;j+1} ~ {\rm for} ~ |j-k| > 1
\nonumber \\
T_{j;j+1} T_{k; k+1} &=& - T_{k;k+1} T_{j;j+1} ~ {\rm for} ~ |j-k| > 1 
\end{eqnarray}
generate the permutation group.  But the candidates $T_{j;j+1} =
(\Gamma_j + \Gamma_{j+1} ) / \sqrt 2$ satisfy the first, second, and 
fourth of these relations.  Thus the
permutation group is realized projectively.

For even $N=2p$ one can construct an irreducible representation of the
$\Gamma$ matrices of dimension $2^p$ iteratively, according to

\begin{eqnarray}
\Gamma_1 &=& \sigma_1 \times 1 \times 1 \times \cdots \nonumber \\
\Gamma_2 &=& \sigma_2 \times 1 \times 1 \times \cdots \nonumber \\
\Gamma_3 &=& \sigma_3 \times \sigma_1 \times 1  \times \cdots\nonumber  \\
\Gamma_4 &=& \sigma_3 \times \sigma_2 \times 1 \times \cdots 
\label{gensAndRels}
\end{eqnarray}
and so forth.  This is not irreducible for $SO(2p)$, because $\kappa
\equiv \Gamma_1 \Gamma_2 \dots \Gamma_{2p}$ commutes with all the
$L_{jk}$.  By projecting onto the eigenvalues of $\kappa$ we get
irreducible spinor representations.  $\kappa$, of course, does not
commute with the representatives of the permutation group.  But
$\tilde \kappa \equiv \kappa (\Gamma_1 - \Gamma_2 + \Gamma_3 -
\Gamma_4 ... ) $
does.  By projecting onto its eigenvalues, we obtain irreducible
(projective) representations of $S_{2p}$.

For odd $N= 2p +1$ we may use the $\Gamma$ matrices constructed for
$N=2p$, together with $\Gamma_{2p+1} = \kappa_{2p}$ to generate the
Clifford algebra.  This gives an irreducible representation for both
the rotation and permutation groups, with no projections necessary.

Schur demonstrated that all the non-trivial, irreducible projective
representations of $S_N$ realize the modified algebra using the first,
second, and fourth relation of ~\ref{gensAndRels}.  Furthermore the
irreducible projective representations may be classified using Young
diagrams, similar to those used for ordinary representations, but with
the additional restriction that row lengths must be strictly
decreasing.  In this construction, the spinorial representation
constructed above corresponds to a single row, analogous to bosons.
There is no corresponding fermion analogue.

In recent work on the Pfaffian $\nu = {1\over 2}$ quantum Hall state,
it was shown that $2n$ quasiparticles at fixed positions span a $2^{n-
1}$ dimensional Hilbert  space, and that braiding such quasiparticles
around one another generated operations closely analogous to spinor
representations \cite{nayak}.  (In addition, there are $e^{2\pi i/8}$
``anyonic'' phase factors.)  The concepts explained above allow one to
formulate the results in a different way: the exchange of these
quasiparticles realizes the simplest projective representation of the
symmetric group.

Another perspective on the projective statistics arises from
realizing the Clifford
algebra in terms of fermion creation and annihilation operators, in
the form
$\Gamma_{2j-1} = e^{-i\pi/4} a_j^* + e^{i\pi/4} a_j$,
$\Gamma_{2j} = e^{i\pi/4} a_j^* + e^{-i\pi/4} a_j$.   Then we find for
the interchange
of an odd index particle with the following even index particle
$(\Gamma_{2j-1} + \Gamma_{2j} )/\sqrt 2  = a_j^* + a_j$, that is,
simply the operation of changing the occupation of the $j^{\rm th}$
mode.  This makes contact with 
an alternative description of the $\nu = 1/2$
quasiparticles using antisymmetric polynomial wave-functions, 
which can be considered
to label occupation numbers of fermionic states \cite{reza}.   
Other exchange operations are a bit more complicated.  In
this notation the operation $\kappa$ is $e^{-i\pi N}$ times a factor
-1 for each unoccupied mode.   Thus projecting to eigenvalues of
$\kappa$ amounts to restricting attention to either even or odd mode
occupations.  This is adequate to
get irreducible representations of the rotation group or of the even
permutations.   If we want to get an irreducible representation  of
all permutations we must allow both even and odd occupations, with a peculiar
global relation between them.

Since the definition of projective statistics refers to interchanges of 
particles, as opposed to braiding, this concept is not in principle tied 
to 2+1 dimensional theories.  Also, no violation of the discrete symmetries 
P, T is implied.

\end{document}